\documentclass[preprintnumbers,amsmath,amssymb,floatfix,12pt,prd,superscriptaddress,nofootinbib]{revtex4}
\usepackage{graphicx}
\usepackage{epsfig}
\usepackage{bm}
\usepackage{amsfonts}

\begin{document}


\title{Polytropic and Chaplygin $f(R)$-gravity models}

\author{K. Karami}
\email{KKarami@uok.ac.ir} \affiliation{Department of Physics,
University of Kurdistan, Pasdaran St., Sanandaj, Iran }
\affiliation{Research Institute for Astronomy $\&$ Astrophysics of
Maragha (RIAAM), Maragha, Iran}

\author{M.S. Khaledian}
\email{MS.Khaledian@uok.ac.ir} \affiliation{Department of Physics,
University of Kurdistan, Pasdaran St., Sanandaj, Iran }

\date{\today}

\begin{abstract}
\vspace*{1.5cm} \centerline{\bf Abstract} \vspace*{0.5cm} We
reconstruct different $f(R)$-gravity models corresponding to the
polytropic, standard Chaplygin, generalized Chaplygin, modified
Chaplygin and modified variable Chaplygin gas dark energy models. We
also obtain the equation of state parameters of the corresponding
$f(R)$-gravity models which describe the accelerated expansion of
the universe. We conclude that although the equation of state
parameters of the obtained $f(R)$-gravities can behave like phantom
or quintessence dark energy models, they cannot justify the
transition from the quintessence state to the phantom regime.
Furthermore, the polytropic and Chaplygin $f(R)$-gravity models
in de Sitter space can satisfy the inflation condition.\\


\end{abstract}

\maketitle

\newpage
\section{Introduction}
Observational data of type Ia supernovae (SNeIa) collected by Riess
et al. \cite{riess} in the High-redshift Supernova Search Team and
by Perlmutter et al. \cite{perl} in the Supernova Cosmology Project
Team independently reported that the present observable universe is
undergoing an accelerated expansion phase. The exotic source for
this cosmic acceleration is generally dubbed ``dark energy'' (DE).
Despite many years of research (see e.g., the reviews
\cite{Padmanabhan,Peebles,Copeland}) its origin has not been
identified till yet. DE is distinguished from ordinary matter (such
as baryons and radiation), in the sense that it has negative
pressure. This negative pressure leads to the accelerated expansion
of the universe by counteracting the gravitational force. The
astrophysical observations show that about 70\% of the present
energy of the universe is contained in DE. There are several DE
models to explain cosmic acceleration.

One of interesting DE models is the polytropic gas which was
introduced by Karami et al. \cite{Karami2} to explain the
accelerated expansion of the universe. The polytropic gas model
plays a very important role in astrophysics. It is still very useful
as a simple example which is nevertheless not too dissimilar from
realistic models. More importantly, there are cases where a
polytropic equation of state is a good approximation to reality. An
example is a gas where the pressure is dominated by degenerate
electrons in white dwarfs or degenerate neutrons in neutron stars.
Another example is the case where pressure and density are related
adiabatically in main sequence stars \cite{Karami2}.

One of another interesting DE candidates is the Chaplygin gas (CG)
which was proposed to explain the accelerated expansion of the
universe \cite{Kamench,Kamenshchik,Bilic}. The simplest type of the
CG models is the standard CG (SCG) which behaves as a dust-like
matter at early times and like a cosmological constant at late stage
\cite{Kamench,Kamenshchik,Bilic}. This interesting feature leads to
the SCG model being proposed as a candidate for the unified DM-DE
(UDME) scenario \cite{Wu,Lu}. The SCG model cannot explain the
astrophysical problems such as structure formation and cosmological
perturbation power spectrum \cite{sandvik,Bean}, hence the
generalized CG (GCG) model was introduced to construct viable
cosmological models \cite{Bento,Gorini,Alam,Bento1,Jamil,Karami}.
After the GCG was introduced, the new model of CG which is called
modified CG (MCG) was proposed. An interesting feature of MCG is
that it can explain the evolution of the universe from radiation to
$\Lambda$CDM \cite{Benaoum,Debnath,Jamil2}. More recently, the
modified variable CG (MVCG) was introduced to describe the evolution
of the universe from radiation era to phantom model
\cite{Anup,Jamil3,Jamil4,Malekjani}. The MVCG can also explain the
flat rotational curves of galaxies \cite{Tekola}.

One of among other interesting alternative proposals for
 DE is modified gravity. It can explain naturally the unification of earlier and
later cosmological epochs (for review see \cite{Capozziello1}).
Moreover, modified gravity may serve as dark matter (DM)
\cite{Sobouti}. There are some classes of modified gravities
containing $f(R)$, $f(\mathcal{G})$, $f(R,\mathcal{G})$ and $f(T)$
which are considered as gravitational alternative for DE
\cite{NojiriOdin,Nojiri1,Nojiri2,Nojiri3,husawiski,Kerner,Barrow,Faraoni,Schmidt,Hu,noj
abdalla,Odintsov,Abdalla,Noj2,Carroll,Capozziello,Nojiri5,Bisabr,Karami4,Khodam-Mohammadi,bengochea}.
Here, $R$ and ${\mathcal
G}=R_{\mu\nu\rho\sigma}R^{\mu\nu\rho\sigma}-4R_{\mu\nu}R^{\mu\nu}+R^2$
are the Ricci scalar and Gauss-Bonnet invariant term, respectively.
Also $R_{\mu\nu\rho\sigma}$ and $R_{\mu\nu}$ are the Riemann and
Ricci tensors, respectively, and $T$ is the torsion scalar.

Note that although the recent observational data from SNeIa,
Wilkinson Microwave Anisotropy Probe (WMAP), Sloan Digital Sky
Survey (SDSS) and so on show remarkably the consistence of the
cosmological constant, it is worth noting that a class of dynamical
models with the equation of state (EoS) parameter across $-1$ dubbed
quintom is mildly favored \cite{YFCai,YFCai1}. In quintom scenario,
there is a no-go theorem which forbids the EoS parameter of a single
perfect fluid or a single scalar field to cross the $-1$ boundary in
the framework of standard Einstein gravity \cite{JQXia}.

Here, our aim is to study how the $f(R)$-gravity can describe the
polytropic, SCG, GCG, MCG and MVCG models as effective theories of
DE models. This motivate us to establish different models of
$f(R)$-gravity according to the above-mentioned scenarios. This
paper is organized as follows. In section 2, we review the theory of
$f(R)$-gravity in the metric formalism. In sections 3 to 8, we
reconstruct different $f(R)$-gravity models corresponding to the
polytropic, SCG, GCG, MCG and MVCG models. Section 9 is devoted to
our conclusions.

\section{$f(R)$-gravity}
The action of $f(R)$-gravity is given by
\cite{NojiriOdin,Nojiri1,Nojiri2,Nojiri3,husawiski,Kerner,Barrow,Faraoni,Schmidt,Hu}
\begin{equation}\label{action}
S = \int {\sqrt { - g} }~{\rm d}^4 x\left[\frac{{R+f(R)}}{{2k^2 }} +
L_{\rm matter}\right],
\end{equation}
where $k^2 =8\pi G$. Also $G$, $g$ and $L_{\rm matter}$ are the
gravitational constant, the determinant of the metric $g_{\mu\nu}$
and the lagrangian density of the matter inside the universe,
respectively.

For the spatially flat Friedmann-Robertson-Walker (FRW) metric, the
Friedmann equations in $f(R)$ theory can be written as
\cite{NojiriOdin2009}
\begin{eqnarray}
&&\frac{3}{k^2}H^2= \rho_m+\rho_R,\label{FiEq1}\\
&&\frac{1}{k^2}(2\dot{H}+3H^2)=-(p_m+p_R),\label{FiEq2}
\end{eqnarray}
where $H=\dot{a}/a$ is the Hubble parameter, $\rho_m$ and $p_m$ are
the energy density and pressure of the matter inside the universe.
Also
\begin{eqnarray}
\rho_{R} =&&\frac{1}{k^2}\left[-\frac{1}{2}f(R)+3\big(\dot{H}+
H^2\big)f'(R)-18\big(4 H^2\dot{H}+
H\ddot{H}\big)f''(R)\right],\label{density}\\
p_{R}=&&\frac{1}{k^2}\Big[\frac{1}{2}f(R)-\big(\dot{H}+3H^2\big)f'(R)
\nonumber\\
&&+6\big(8H^2\dot{H}+6H\ddot{H}+4{\dot{H}}^2+\dddot{H}\big)f''(R)+36
\big(\ddot{H}+4H\dot{H}\big)^2f'''(R)\Big],\label{pressure}
\end{eqnarray}
with
\begin{equation}
R =6(\dot{H}+2H^2).\label{R}
\end{equation}
Here, the dot and the prime denote the derivatives with respect to
the cosmic time $t$ and $R$, respectively. Also $\rho_{R}$ and
$p_{R}$ are the curvature contribution to the energy density and
pressure.

The energy conservation laws are given by
\begin{eqnarray}
&&\dot{\rho}_{m}+3H(\rho_{m}+p_{m})=0,\\
&&\dot{\rho}_{R}+3H(\rho_{R}+p_{R})=0.~\label{ecT}
\end{eqnarray}
The EoS parameter due to the curvature contribution is defined as
\cite{Nozari}
\begin{eqnarray}
\omega_{R}&=&\frac{p_R}{\rho_R}\nonumber\\
&=&-1-
\frac{4\Big[\dot{H}f'(R)+3\big(3H\ddot{H}-4H^2\dot{H}+4{\dot{H}}^2+\dddot{H}\big)f''(R)
+18\big(\ddot{H}+4H\dot{H}\big)^2f'''(R)\Big]}
{\Big[f(R)-6\big(\dot{H}+ H^2\big)f'(R)+36\big(4 H^2\dot{H}+
H\ddot{H}\big)f''(R)\Big]}.~\label{wHDETotal}
\end{eqnarray}
In what follows, we reconstruct different $f(R)$-gravities according
to the polytropic, SCG, GCG, MCG and MVCG models.
\section{Polytropic $f(R)$-gravity model}

Here, we reconstruct the $f(R)$-gravity from the polytropic gas DE
model. Following \cite{Karami2}, the EoS of the polytropic gas is
given by
\begin{equation}\label{pol1}
p_{\Lambda}=-K\rho_{\Lambda}^{1+\frac{1}{n}},
\end{equation}
where $K$ is a positive constant and $n>0$ is the polytropic index.
Using Eq. (\ref{ecT}) the energy density of the polytropic gas
evolves as
\begin{equation}\label{pol2}
\rho_{\Lambda}=\left(Ca^\frac{3}{n}+K\right)^{-n},
\end{equation}
where $C$ is an integration constant \cite{Karami2}.

Here, we assume an ansatz for the scale factor as
\begin{equation}
a(t)=a_0(t_s-t)^{-h},~~~t\leq t_s,~~~h>0,\label{a}
\end{equation}
which usually people consider for describing the present
accelerating expansion phase of the universe in different modified
gravities like $f(R)$, $f(\mathcal{G})$ and $f(R,\mathcal{G})$
\cite{Nojiri,Sadjadi}. Using Eqs. (\ref{R}) and (\ref{a}) one can
obtain
\begin{equation}
H=\frac{h}{t_s-t}=\left[\frac{h}{6(2h+1)}R\right]^{1/2},~~~\dot{H}=H^2/h.\label{respect
to r}
\end{equation}
Using Eqs. (\ref{a}) and (\ref{respect to r}) one can rewrite the
energy density (\ref{pol2}) in terms of $R$ as
\begin{equation}
\rho_{\Lambda}=\left[C{\left(6h(2h+1){a_0}^{\frac{-2}{h}}\right)}^{\frac{-3h}{2n}}~R^\frac{3h}{2n}+K\right]^{-n}.
\label{pol22}
\end{equation}
Substituting Eq. (\ref{pol22}) into (\ref{density}), i.e.
$\rho_R=\rho_{\Lambda}$, gives
\begin{equation}
2R^2 f''(R)-(h+1)R f'(R)+(2h+1)f(R)+\Big(\varepsilon+\xi
R^{\frac{3h}{2n}}\Big)^{-n}=0,\label{dif eq2}
\end{equation}
where
\begin{eqnarray}
\varepsilon&=&K\Big(2(2h+1)k^2\Big)^{\frac{-1}{n}},\\
\xi&=&\varepsilon\frac{C}{K}{\Big(6h(2h+1){a_0}^{\frac{-2}{h}}\Big)}^{\frac{-3h}{2n}}.\label{TetaFiPoly}
\end{eqnarray}
Solving Eq. (\ref{dif eq2}) yields the polytropic $f(R)$-gravity as
\begin{eqnarray}
f(R)&=&\lambda_+R^{m_+}+\lambda_-R^{m_-}-\frac{1}{\varepsilon^n
(2h+1)(m_+-m_-)}\nonumber\\
&&\times\left[m_+~{_2}F_1\left(\frac{-2nm_-}{3h},n;1-\frac{2nm_-}{3h};
\frac{-\xi R^{\frac{3h}{2n}}}{\varepsilon}\right)-\right.
\nonumber\\
&&~\left.m_-~{_2}F_1\left(\frac{-2nm_+}{3h},n;1-\frac{2nm_+}{3h};
\frac{-\xi
R^{\frac{3h}{2n}}}{\varepsilon}\right)\right],~~~\label{fR-poly}
\end{eqnarray}
where $_{2}F_1$ denotes the first hypergeometric function and
\begin{equation}
m_\pm=\frac{3+h\pm\sqrt{h^2-10h+1}}{4}.\label{alpha}
\end{equation}
Also $\lambda_\pm$ are the integration constants that can be
determined from the necessary boundary conditions. Following
\cite{NojiriOdin1,Nojiri8} the accelerating expansion in the present
universe could be generated, if one consider that $f(R)$ could be a
small constant at present universe, that is
\begin{eqnarray}
&&f(R_0)=-2R_0,\label{bc1}\\
&&f'(R_0)\sim0,\label{bc2}
\end{eqnarray}
where $R_0\sim(10^{-33}{\rm eV})^2$ is the current curvature.
Applying the above boundary conditions to the solution
(\ref{fR-poly}) one can obtain
\begin{eqnarray}
\lambda_{+}&=&\frac{m_-}{(2h+1)(m_+-m_-)R_0^{m_+}\alpha^{n}}
\nonumber\\
&&\times\left[2(2h+1)R_0\alpha^{n}-{_2}F_1\left(\frac{-2nm_+}{3h},n;1-\frac{2nm_+}{3h};
\frac{-\xi R_0^{\frac{3h}{2n}}}{\varepsilon}\right)\right],
\end{eqnarray}
\begin{eqnarray}
\lambda_{-}&=&\frac{m_+}{(2h+1)(m_--m_+)R_0^{m_-}\alpha^{n}}
\nonumber\\
&&\times\left[2(2h+1)R_0\alpha^{n}-{_2}F_1\left(\frac{-2nm_-}{3h},n;1-\frac{2nm_-}{3h};
\frac{-\xi R_0^{\frac{3h}{2n}}}{\varepsilon}\right)\right].
\end{eqnarray}
Inserting Eq. (\ref{fR-poly}) into (\ref{wHDETotal}) and using
(\ref{respect to r}) one can get the EoS parameter of the polytropic
$f(R)$-gravity model as
\begin{equation}\label{wpoly1}
\omega_{R} =-1+\frac{1}{1+\frac{\varepsilon}{\xi}R^{\frac{-3h}{2n}}}
=-1+\frac{1}{1+\frac{K}{C}a^{\frac{-3}{n}}},
\end{equation}
which can be also obtained from
$\omega_{\Lambda}=p_{\Lambda}/\rho_{\Lambda}=-K\rho_{\Lambda}^{\frac{1}{n}}$.
Using the redshift $z =\frac{1}{a}-1$, where we take $a_0=1$ for the
present value of the scale factor, the above relation can be
rewritten as
\begin{equation}\label{wCGDEz}
\omega_{R}=-1+\frac{1}{1+\frac{K}{C}\left[1+z\right]^{\frac{3}{n}}}.
\end{equation}
We see that for $C>0$ then $\omega_{R}>-1$ which corresponds to a
quintessence-like accelerating universe. For $C<0$, one can rewrite
Eqs. (\ref{pol2}) and (\ref{wCGDEz}) as
\begin{eqnarray}
\rho_{\Lambda}&=&K^{-n}\left[1-\left(\frac{1+z_{\rm crit}}{1+z}\right)^{\frac{3}{n}}\right]^{-n},\label{pol3}\\
\omega_{R}&=&-1+\frac{1}{1-\left[\frac{1+z}{1+z_{\rm
crit}}\right]^{\frac{3}{n}}},\label{wCGDEzc1}
\end{eqnarray}
where
\begin{equation}
z_{\rm crit}=\left(\frac{|C|}{K}\right)^{\frac{n}{3}}-1,
\end{equation}
is the critical redshift. Equation (\ref{wCGDEzc1}) shows that for
$z<z_{\rm crit}$ the EoS parameter behaves like quintessence DE
($\omega_R>-1$). Here due to having $\rho_{\Lambda}>0$, from Eq.
(\ref{pol3}) we see that the polytropic index should be even,
$n=(2,4,6,\cdot\cdot\cdot)$. For $z>z_{\rm crit}$, from Eq.
(\ref{wCGDEzc1}) we have $\omega_R<-1$ which behaves as a phantom
type DE. Note that here crossing the phantom-divide line cannot
occur because at $z=z_{\rm crit}$ we have $\omega_R\rightarrow
\infty$. In other words, the EoS parameter (\ref{wCGDEz}) cannot
justify the transition from the quintessence state, $\omega_R>-1$,
to the phantom regime, $\omega_R<-1$, as indicated by recent
observations \cite{Sahni,Huterer5,Wang6}.
\section{Standard Chaplygin $f(R)$-gravity model}

The EoS of the SCG model of DE is as follows
\cite{Kamench,Kamenshchik,Bilic}
\begin{equation}
p_{\Lambda}=-\frac{A}{\rho_{\Lambda}},\label{SCGEoS}
\end{equation}
where $A$ is a positive constant. Inserting the above EoS into the
energy conservation equation (\ref{ecT}), leads to a density
evolving as \cite{Kamench,Kamenshchik,Bilic}
\begin{equation}\label{SCG}
\rho_{\Lambda}=\sqrt{A+\frac{C}{a^6}},
\end{equation}
where $C$ is an integration constant. Note that the SCG model offers
a unified picture of DM and DE \cite{Wu,Lu}. Because it smoothly
interpolates between a non-relativistic matter phase
($\rho_{\Lambda}\propto a^{-3}$) in the past and a negative-pressure
DE regime ($\rho_{\Lambda}=-p_{\Lambda}$) at late times.

Using Eqs. (\ref{a}) and (\ref{respect to r}) the energy density
(\ref{SCG}) can be rewritten as
\begin{equation}\label{rhoSCG}
\rho_{\Lambda}=\sqrt{A+C{\left(6h(2h+1){a_0}^{\frac{-2}{h}}\right)}^{3h}
R^{-3h}}.
\end{equation}
Equating (\ref{rhoSCG}) with (\ref{density}), i.e.
$\rho_R=\rho_{\Lambda}$, gives
\begin{equation}\label{difeqSCG}
2R^2 f''(R)-(h+1)R
f'(R)+(2h+1)f(R)+\sqrt{\varepsilon+\frac{\xi}{R^{3h}}}=0,
\end{equation}
where
\begin{eqnarray}\label{TetaFiSCG}
\varepsilon&=&A\Big(2(2h+1)k^2\Big)^2,\\
\xi&=&\varepsilon \frac{C}{A}{\left(6h(2h+1){a_0}^{\frac{-2}{h}}\right)}^{3h}.
\end{eqnarray}
Solving the differential Eq. (\ref{difeqSCG}) yields
\begin{eqnarray}\label{fR-SCG1}
f(R)=\lambda_+R^{m_+}+\lambda_-R^{m_-}&-&\frac{\sqrt{\varepsilon}}{(2h+1)(m_+-m_-)}
\nonumber\\
&\times&\left[m_+~{_2}F_1\left(\frac{m_-}{3h},\frac{-1}{2};1+\frac{m_-}{3h};
\frac{-\xi}{\varepsilon R^{3h}} \right)-\right.
\nonumber\\
&&\left.m_-~{_2}F_1\left(\frac{m_+}{3h},\frac{-1}{2};1+\frac{m_+}{3h};
\frac{-\xi}{\varepsilon R^{3h}}\right) \right],
\end{eqnarray}
where $m_{\pm}$ are given by Eq. (\ref{alpha}). Also $\lambda_\pm$
are determined from the boundary conditions (\ref{bc1}) and
(\ref{bc2}) as
\begin{eqnarray}
\lambda_{+}&=&\frac{m_-}{(2h+1)(m_+-m_-)R_0^{m_+}}\nonumber\\&&\times
\left[2(2h+1)R_0-\sqrt{\varepsilon}~{_2}F_1\left(\frac{m_+}{3h},-\frac{1}{2};1+\frac{m_+}{3h};
\frac{-\xi}{\varepsilon R_0^{3h}}\right)\right],
\end{eqnarray}
\begin{eqnarray}
\lambda_{-}&=&\frac{m_+}{(2h+1)(m_--m_+)R_0^{m_-}}\nonumber\\&&\times
\left[2(2h+1)R_0-\sqrt{\varepsilon}~{_2}F_1\left(\frac{m_-}{3h},-\frac{1}{2};1+\frac{m_-}{3h};
\frac{-\xi}{\varepsilon R_0^{3h}}\right)\right].
\end{eqnarray}
Substituting Eq. (\ref{fR-SCG1}) into (\ref{wHDETotal}) and using
(\ref{respect to r}) one can get the EoS parameter of the standard
Chaplygin $f(R)$-gravity model as
\begin{equation}\label{wSCG1}
\omega_{R}= -1+\frac{1}{1+\frac{\varepsilon}{\xi}R^{3h}}=
-1+\frac{1}{1+\frac{A}{C}a^{6}},
\end{equation}
which is same as
$\omega_{\Lambda}=p_{\Lambda}/\rho_{\Lambda}=-A\rho_{\Lambda}^{-2}$.
The above relation can be rewritten in terms of redshift as
\begin{equation}
\omega_{R}=-1+\frac{1}{1+\frac{A}{C}\left[1+z\right]^{-6}}.\label{wSCGz}
\end{equation}
We see that for $C>0$, $\omega_{R}>-1$ which behaves like
quintessence accelerating universe. For $C<0$, rewriting Eqs.
(\ref{SCG}) and (\ref{wSCGz}) give
\begin{eqnarray}
\rho_{\Lambda}&=&\left\{A\left[1-\Big(\frac{1+z}{1+z_{\rm
crit}}\Big)^{6}\right]\right\}^{1/2},\label{pol4}\\
\omega_{R}&=&-1+\frac{1}{1-\left[\frac{1+z_{\rm crit}}{1+z}\right]^6},\label{wSCGzc}
\end{eqnarray}
where
\begin{equation}\label{ZcrSCG}
   z_{\rm crit}=\left(\frac{A}{|C|}\right)^{\frac{1}{6}}-1.
\end{equation}
Equation (\ref{pol4}) shows that due to having a positive energy
density, the redshift parameter should be in the range of $z<z_{\rm
crit}$. Now for $z<z_{\rm crit}$, Eq. (\ref{wSCGzc}) clears that
$\omega_R<-1$ which corresponds to a universe dominated by phantom
type DE. Therefore, the EoS parameter (\ref{wSCGz}) like the
polytropic model (\ref{wCGDEz}) cannot realize the quintom behavior,
with an EoS crossing $-1$. Indeed, the parameterization of energy
density such as Eq. (\ref{SCG}) doesn't yield the $-1$-crossing
during the evolution of the universe. In \cite{YFCai2}, it was shown
that a unified model of quintom and Chaplygin gas can be constructed
by virtue of a nonconventional spinor field explicitly.

\section{Generalized Chaplygin $f(R)$-gravity model}

The EoS of the GCG model of DE takes the form
\cite{Bento,Gorini,Alam,Bento1,Jamil,Karami}
\begin{equation}
p_{\Lambda}=-\frac{A}{\rho_{\Lambda}^\alpha},\label{GCGEoS}
\end{equation}
where $A$ a positive constant and $\alpha$ is a constant in the
range of $0\leq\alpha\leq 1$ (the SCG corresponds to the case
$\alpha=1$). Using Eq. (\ref{ecT}), the GCG energy density evolves
as \cite{Bento,Gorini,Alam,Bento1,Jamil,Karami}
\begin{equation}
\rho_{\Lambda}=\left({A+\frac{C}{a^{3(1+\alpha)}}}\right)^{\frac{1}{1+\alpha}},\label{GCG}
\end{equation}
where $C$ is an integration constant.

Using Eqs. (\ref{a}) and (\ref{respect to r}) the energy density
(\ref{GCG}) takes the form
\begin{equation}
\rho_{\Lambda}=\Big[A+C{\Big(6h(2h+1){a_0}^{\frac{-2}{h}}\Big)}^{\frac{3h(1+\alpha)}{2}}
R^{\frac{-3h(1+\alpha)}{2}}\Big]^{\frac{1}{1+\alpha}}.\label{rhoGCG}
\end{equation}
Equating (\ref{rhoGCG}) with (\ref{density}), one can obtain
\begin{equation}
2R^2 f''(R)-(h+1)R
f'(R)+(2h+1)f(R)+\Big[\varepsilon+\frac{\xi}{R^{\frac{3h\gamma}{2}}}\Big]^{\frac{1}
{\gamma}}=0,\label{dif eqGCG}
\end{equation}
where
\begin{eqnarray}
\varepsilon&=&A~\Big(2(2h+1)k^2\Big)^\gamma,\\
\xi&=&\varepsilon\frac{C}{A}{\Big(6h(2h+1){a_0}^{\frac{-2}{h}}\Big)}^{\frac{3h\gamma}{2}},\\
\gamma&=&1+\alpha.\label{TeatFiGCG}
\end{eqnarray}
Solving the differential Eq. (\ref{dif eqGCG}) yields
\begin{eqnarray}
f(R)=\lambda_+R^{m_+}+\lambda_-R^{m_-}&-&\frac{{\varepsilon}^{\frac{1}{\gamma}}}
{(2h+1)(m_+-m_-)}\nonumber\\
&\times&\left[m_+~{_2}F_1\left(\frac{2m_-}{3h\gamma},\frac{-1}{\gamma};1+\frac{2m_-}{3h\gamma};
\frac{-\xi}{\varepsilon R^{\frac{3h\gamma}{2}}}\right)-\right.
\nonumber\\
&&\left.m_-~{_2}F_1\left(\frac{2m_+}{3h\gamma},\frac{-1}{\gamma};1+\frac{2m_+}{3h\gamma};
\frac{-\xi}{\varepsilon R^{\frac{3h\gamma}{2}}} \right)
\right],\label{fR-GCG}
\end{eqnarray}
where $m_{\pm}$ are given by Eq. (\ref{alpha}). Also $\lambda_\pm$
are determined by applying the boundary conditions (\ref{bc1}) and
(\ref{bc2}) to the solution (\ref{fR-GCG}). The resulting
$\lambda_\pm$ are
\begin{eqnarray}
\lambda_{+}&=&\frac{m_-}{(2h+1)(m_+-m_-)R_0^{m_+}}\nonumber\\
&&\times\left[2(2h+1)R_0-{\varepsilon}^\frac{1}{\gamma}~
{_2}F_1\left(\frac{2m_+}{3h\gamma},\frac{-1}{\gamma};1+\frac{2m_+}{3h\gamma};
\frac{-\xi}{\varepsilon R_0^{\frac{3h\gamma}{2}}}\right)\right],
\end{eqnarray}
\begin{eqnarray}
\lambda_{-}&=&\frac{m_+}{(2h+1)(m_--m_+)R_0^{m_-}}\nonumber\\
&&\times\left[2(2h+1)R_0-{\varepsilon}^\frac{1}{\gamma}~
{_2}F_1\left(\frac{2m_-}{3h\gamma},\frac{-1}{\gamma};1+\frac{2m_-}{3h\gamma};
\frac{-\xi}{\varepsilon R_0^{\frac{3h\gamma}{2}}}\right)\right].
\end{eqnarray}
Inserting Eq. (\ref{fR-GCG}) into (\ref{wHDETotal}) and using
(\ref{respect to r}) one can get the EoS parameter of the
generalized Chaplygin $f(R)$-gravity model as
\begin{equation}\label{wGCG1}
\omega_{R}= -1+\frac{1}{1+\frac{\varepsilon}{\xi}
R^{\frac{3h\gamma}{2}}}=-1+\frac{1}{1+\frac{A}{C}a^{3(1+\alpha)}},
\end{equation}
which is same as $\omega_{\Lambda}=p_{\Lambda}/\rho_{\Lambda}=-{A}/\
\rho_{\Lambda}^{1+\alpha}$. Using the redshift we get
\begin{equation}\label{wGCGz}
\omega_{R}=-1+\frac{1}{1+\frac{A}{C}\left[1+z\right]^{-3(1+\alpha)}}.
\end{equation}
The above relation clears that for $C>0$, $\omega_{R}>-1$ which acts
as a quintessence DE. For $C<0$, one can rewrite Eqs. (\ref{GCG})
and (\ref{wGCGz}) as
\begin{eqnarray}
\rho_{\Lambda}&=&A^{\frac{1}{(1+\alpha)}}\left[1-\Big(\frac{1+z}
{1+z_{\rm crit}}\Big)^{3(1+\alpha)}\right]^{\frac{1}{(1+\alpha)}},\label{pol5}\\
\omega_{R}&=&-1+\frac{1}{1-\left(\frac{1+z_{\rm
crit}}{1+z}\right)^{3(1+\alpha)}},\label{wGCGzc}
\end{eqnarray}
where
\begin{equation}
z_{\rm crit}=\left(\frac{A}{|B|}\right)^{\frac{1}{3(1+\alpha)}}-1.
\end{equation}
Equation (\ref{wGCGzc}) shows that for $z<z_{\rm crit}$ and
$z>z_{\rm crit}$ we have the phantom-like ($\omega_R<-1$) and
quintessence-like ($\omega_R>-1$) accelerating universes,
respectively. Note that here crossing the phantom-divide line cannot
occur because the critical redshift separates the phantom region
($z<z_{\rm crit}$) from the quintessence regime ($z>z_{\rm crit}$).
Furthermore, for $z>z_{\rm crit}$ due to having $\rho_{\Lambda}>0$,
from Eq. (\ref{pol5}) we need to have $\frac{1}{1+\alpha}=2Q$ where
$Q$ is a fractional number in the range of $1/4<Q<1/2$ and its
denominator should be odd.
\section{Modified Chaplygin $f(R)$-gravity model}

The EoS of the MCG model of DE is given by
\cite{Benaoum,Debnath,Jamil2}
\begin{equation}
p_{\Lambda}=B\rho_\Lambda-\frac{A}{\rho_{\Lambda}^\alpha},\label{MCGEoS}
\end{equation}
where $A$ and $B$ are positive constants and $0\leq\alpha\leq1$.
Using Eq. (\ref{ecT}), the MCG energy density evolves as
\cite{Benaoum,Debnath,Jamil2}
\begin{equation}\label{MCG}
\rho_{\Lambda}=\left({\frac{A}{1+B}+\frac{C}{a^{3(1+\alpha)(1+B)}}}\right)^{\frac{1}{1+\alpha}},
\end{equation}
where $C$ is an integration constant. Using Eqs. (\ref{a}) and
(\ref{respect to r}) the above relation yields
\begin{equation}\label{rhoCG}
\rho_{\Lambda}=\left[\frac{A}{1+B}+C{\left(6h(2h+1){a_0}^{\frac{-2}{h}}\right)}
^{\frac{3h(1+\alpha)(1+B)}{2}}R^{\frac{-3h(1+\alpha)(1+B)}{2}}\right]^{\frac{1}{1+\alpha}}.
\end{equation}
Equating (\ref{rhoCG}) with (\ref{density}) gives
\begin{equation}\label{dif eq6}
2R^2 f''(R)-(h+1)R
f'(R)+(2h+1)f(R)+\Big[\varepsilon+\frac{\xi}{R^{\frac{3h\gamma\eta}{2}}}\Big]^
{\frac{1}{\gamma}}=0,
\end{equation}
where
\begin{eqnarray}\label{TeatFiMCG}
\varepsilon&=&A(1+B)^{-1}\Big(2(2h+1)k^2\Big)^\gamma ,\\
\xi&=&C~\left(6h(2h+1){a_0}^{\frac{-2}{h}}\right)^{\frac{3h\gamma\eta}{2}}\Big(2(2h+1)k^2\Big)^\gamma,\\
\gamma&=&1+\alpha,\\
\eta&=&1+B.
\end{eqnarray}
Solving Eq. (\ref{dif eq6}) gets
\begin{eqnarray}\label{fR-MCG}
f(R)=\lambda_+R^{m_+}+\lambda_-R^{m_-}&-&\frac{{\varepsilon}^{\frac{1}{\gamma}}}
{(2h+1)(m_+-m_-)}\nonumber\\
&\times&\left[m_+~{_2}F_1\left(\frac{2m_-}{3h\gamma\eta},\frac{-1}{\gamma};1+\frac{2m_-}{3h\gamma\eta};
\frac{-\xi}{\varepsilon R^{\frac{3h\gamma\eta}{2}}} \right)-\right.
\nonumber\\
&&\left.m_-~{_2}F_1\left(\frac{2m_+}{3h\gamma\eta},\frac{-1}{\gamma};1+\frac{2m_+}{3h\gamma\eta};
\frac{-\xi}{\varepsilon R^{\frac{3h\gamma\eta}{2}}} \right)\right],
\end{eqnarray}
where $m_{\pm}$ are given by Eq. (\ref{alpha}) and $\lambda_\pm$ are
determined from the boundary conditions (\ref{bc1}) and (\ref{bc2})
as
\begin{eqnarray}
\lambda_{+}&=&\frac{m_-}{(2h+1)(m_+-m_-)R_0^{m_+}}\nonumber\\
&&\times\left[2(2h+1)R_0-{\varepsilon}^\frac{1}{\gamma}~
{_2}F_1\left(\frac{2m_+}{3h\gamma\eta},\frac{-1}{\gamma};1+\frac{2m_+}{3h\gamma\eta};
\frac{-\xi}{\varepsilon R_0^{\frac{3h\gamma\eta}{2}}}\right)\right],
\end{eqnarray}
\begin{eqnarray}
\lambda_{-}&=&\frac{m_+}{(2h+1)(m_--m_+)R_0^{m_-}}\nonumber\\
&&\times\left[2(2h+1)R_0-{\varepsilon}^\frac{1}{\gamma}~
{_2}F_1\Big(\frac{2m_-}{3h\gamma\eta},\frac{-1}{\gamma};1+\frac{2m_-}{3h\gamma\eta};
\frac{-\xi}{\varepsilon R_0^{\frac{3h\gamma\eta}{2}}}\Big)\right].
\end{eqnarray}
Inserting Eq. (\ref{fR-MCG}) into (\ref{wHDETotal}) and using
(\ref{respect to r}) one can get the EoS parameter of the modified
Chaplygin $f(R)$-gravity model as
\begin{eqnarray}\label{wMCG1}
\omega_{R}=-1+\frac{1+B} {1+\frac{\varepsilon}{\xi}
R^{\frac{3h\gamma\eta}{2}}}=-1+\frac{1+B}{1+\frac{A}{C(1+B)}a^{3(1+\alpha)(1+B)}},
\end{eqnarray}
which is same as
$\omega_{\Lambda}=p_{\Lambda}/\rho_{\Lambda}=B-{A}/\
\rho_{\Lambda}^{1+\alpha}$. In terms of redshift we have
\begin{equation}\label{wMCGz}
\omega_{R}=-1+\frac{1+B}{1+\frac{A}{C(1+B)}\left[1+z\right]^{-3(1+\alpha)(1+B)}}.
\end{equation}
The above relation illustrates that for $C>0$, $\omega_{R}>-1$ which
shows a quintessence type of DE. For $C<0$ one can rewrite Eqs.
(\ref{MCG}) and (\ref{wMCGz}) as
\begin{eqnarray}
\rho_{\Lambda}&=&\left\{\frac{A}{1+B}\left[1-\Big(\frac{1+z}{1+z_{\rm
crit}}\Big)^{3(1+\alpha)(1+B)}\right] \right\}^{\frac{1}{1+\alpha}},\label{pol7}\\
\omega_{R}&=&-1+\frac{1}{1-\left[\frac{1+z_{\rm
crit}}{1+z}\right]^{3(1+\alpha)(1+B)}},\label{wMCGzc}
\end{eqnarray}
where
\begin{equation}\label{ZcrMCG}
z_{\rm
crit}=\left(\frac{A}{|C|(1+B)}\right)^{\frac{1}{3(1+\alpha)(1+B)}}-1.
\end{equation}
Equation (\ref{wMCGzc}) shows that for $z<z_{\rm crit}$ and
$z>z_{\rm crit}$, the universe take places in the phantom
($\omega_R<-1$) and quintessence ($\omega_R>-1$) regimes,
respectively. For $z>z_{\rm crit}$ due to having $\rho_{\Lambda}>0$,
from Eq. (\ref{pol7}) we need to have $\frac{1}{1+\alpha}=2Q$ where
$Q$ is a fractional number in the range of $1/4<Q<1/2$ and its
denominator should be odd. Note that in the modified Chaplygin
$f(R)$-gravity model like the previous models, the transition from
the quintessence state to the phantom regime cannot occur.
\section{Modified variable Chaplygin $f(R)$-gravity model}
The MVCG model of DE has the following EoS
\cite{Anup,Jamil3,Jamil4,Malekjani}
\begin{equation}
p_{\Lambda}=B\rho_\Lambda-\frac{A}{a^n\rho_{\Lambda}^\alpha},\label{MVCGEoS}
\end{equation}
where $A$, $B$ and $n$ are positive constants and
$0\leq\alpha\leq1$. Using Eq. (\ref{ecT}), the MVCG energy density
evolves as \cite{Benaoum,Debnath,Jamil2}
\begin{equation}
\rho_{\Lambda}=\frac{1}{a^{\frac{n}{1+\alpha}}}\left({\frac{3(1+\alpha)A}
{3(1+\alpha)(1+B)-n}+\frac{C}{a^{3(1+\alpha)(1+B)-n}}}\right)^{\frac{1}{1+\alpha}},\label{MVCG}
\end{equation}
where $C$ is an integration constant.

Using Eqs. (\ref{a}) and (\ref{respect to r}) one can rewrite
(\ref{MVCG}) as
\begin{equation}
\rho_{\Lambda}=\chi_1~R^{\frac{-nh}{2(1+\alpha)}}\left[\chi_2+\chi_3~R^{\frac{h}{2}\big(n-3(1+\alpha)
(1+B)\big)}\right]^{\frac{1}{1+\alpha}},\label{rhoMVCG}
\end{equation}
where
\begin{eqnarray}
\chi_1&=&{\Big[6h(2h+1){a_0}^{\frac{-2}{h}}\Big]}^{\frac{nh}{2(1+\alpha)}},\\
\chi_2&=&\frac{3(1+\alpha)A}{3(1+\alpha)(1+B)-n},\\
\chi_3&=&C{\Big[6h(2h+1){a_0}^{\frac{-2}{h}}\Big]}^{\frac{h}{2}
\big(3(1+\alpha)(1+B)-n\big)}.\label{alphaMVCG}
\end{eqnarray}
Equating (\ref{rhoMVCG}) with (\ref{density}) gives
\begin{equation}
2R^2
f''(R)-(h+1)R f'(R)+(2h+1)f(R)+\frac{1}{R^{\sigma}}\Big[\varepsilon+\frac{\xi}{R^{\delta}}\Big]^
{\frac{1}{\gamma}}=0,\label{dif eqMVCG}
\end{equation}
where
\begin{eqnarray}
\varepsilon&=&\chi_2~\chi_1^{\gamma}
\big[2(2h+1)k^2\big]^\gamma,\\
\xi&=&C\Big(\frac{\varepsilon~\chi_3}{\chi_2}\Big),\\
\gamma&=&1+\alpha,\\
\delta&=&\frac{3h(1+\alpha)A}{2\chi_2},\\
\sigma&=&\frac{nh}{2(1+\alpha)}.\label{TeatFiMVCG}
\end{eqnarray}

Solving the differential Eq. (\ref{dif eqMVCG}) yields
\begin{eqnarray}\label{fR-MVCG}
f(R)=\lambda_+R^{m_+}+\lambda_-R^{m_-}&-&\frac{{\varepsilon}^{\frac{1}{\gamma}}}
{\big(h(2+\sigma)+(1+\sigma)(1+2\sigma)\big)(m_+-m_-)}\nonumber\\
&\times&\left[(m_++\sigma)~~{_2}F_1\left(\frac{m_-+\sigma}{\delta},\frac{-1}{\gamma};1+\frac{m_-+\sigma}{\delta};
\frac{-\xi}{\varepsilon R^{\delta}} \right)-\right.
\nonumber\\
&&\left.(m_-+\sigma)~~{_2}F_1\left(\frac{m_++\sigma}{\delta},\frac{-1}{\gamma};1+\frac{m_++\sigma}{\delta};
\frac{-\xi}{\varepsilon R^{\delta}}\right) \right],~
\end{eqnarray}
where $m_{\pm}$ are given by Eq. (\ref{alpha}) and $\lambda_\pm$ are
determined from the boundary conditions (\ref{bc1}) and (\ref{bc2})
as
\begin{eqnarray}
\lambda_{+}=\frac{1}{4(m_+-m_-)R_0^{m_++\sigma}}\left[8m_-R_0^{1+\sigma}+
\frac{4(m_-+\sigma)\varepsilon^{\frac{1}{\gamma}}}{h(2+\sigma)
+(1+\sigma)(1+2\sigma)}\right.~~~~~~~~~~~~~~~~~\nonumber\\
\left.\times{_2}F_1\left(\frac{m_++\sigma}{\delta},\frac{-1}{\gamma};1+\frac{m_++\sigma}{\delta};
\frac{-\xi}{\varepsilon R_0^\delta}\right)\right],
\end{eqnarray}
\begin{eqnarray}
\lambda_{-}=\frac{1}{4(m_--m_+)R_0^{m_-+\sigma}}\left[8m_+R_0^{1+\sigma}+
\frac{4(m_++\sigma)\varepsilon^{\frac{1}{\gamma}}}{h(2+\sigma)
+(1+\sigma)(1+2\sigma)}\right.~~~~~~~~~~~~~~~~~\nonumber\\
\left.\times{_2}F_1\left(\frac{m_-+\sigma}{\delta},\frac{-1}{\gamma};1+\frac{m_-+\sigma}{\delta};
\frac{-\xi}{\varepsilon R_0^\delta}\right)\right].
\end{eqnarray}
Substituting Eq. (\ref{fR-MVCG}) into (\ref{wHDETotal}) and using
(\ref{respect to r}) yields the EoS parameter of the modified
variable Chaplygin $f(R)$-gravity model as
\begin{eqnarray}
\omega_{R}&=&-1+\frac{1}{3(1+\alpha)}\left[n+\frac{\beta}{1+
\frac{3(1+\alpha)A}{C\beta}\Big(\frac{R}{6h(2h+1)}\Big)^{\frac{h\beta}{2}}}\right],\nonumber\\
&=&-1+\frac{1}{3(1+\alpha)}\left[n+\frac{\beta}{1+
\frac{3(1+\alpha)A}{C\beta}~a^{\beta}}\right], \label{wMVCGz1}
\end{eqnarray}
where
\begin{equation}\label{beta}
    \beta=3(1+\alpha)(1+A)-n.
\end{equation}
This is same as
$\omega_{\Lambda}=p_{\Lambda}/\rho_{\Lambda}=B-{A}/(a^n
\rho_{\Lambda}^{1+\alpha})$. In terms of redshift it yields
\begin{equation}
\omega_{R}=-1+\frac{1}{3(1+\alpha)}\left[n+\frac{\beta}{1+\frac{3(1+\alpha)A}{C\beta}(1+z)^{-\beta}}\right].
\label{wMVCGz}
\end{equation}
Equation (\ref{wMVCGz}) shows that for $C>0$ we have a
quintessence-like EoS parameter, $\omega_R>-1$. For $C<0$ there is a
critical redshift when $\omega_{R}\rightarrow\infty$. For this case
we rewrite Eqs. (\ref{MVCG}) and (\ref{wMVCGz}) as
\begin{eqnarray}
\rho_{\Lambda}&=&\left[\frac{3A(1+\alpha)(1+z)^n}{\beta}
\left(1-\Big(\frac{1+z}{1+z_{\rm
crit}}\Big)^{\beta}\right)\right]^{\frac{1}{1+\alpha}},\label{pol6}\\
\omega_{R}&=&-1+\frac{1}{3(1+\alpha)}\left(n+\frac{\beta}{1-\left(\frac{1+z_{\rm
crit}}{1+z}\right)^{\beta}}\right)~,\label{wMVCGzc}
\end{eqnarray}
where
\begin{equation}
    z_{\rm crit}=\left(\frac{3A(1+\alpha)}{C\beta}\right)^{\frac{1}{\beta}}-1.
\end{equation}
Equation (\ref{wMVCGzc}) clears that the EoS parameter for $z<z_{\rm
crit}$ and $z>z_{\rm crit}$ behaves like the phantom ($\omega_R<-1$)
and quintessence ($\omega_R>-1$) DE models. The result is the same
as that obtained for the EoS parameter of the modified Chaplygin
$f(R)$-gravity model (\ref{wMCGzc}) and independent of the sign of
parameter $\beta$. Besides, for $z>z_{\rm crit}$ due to having
$\rho_{\Lambda}>0$, from Eq. (\ref{pol6}) we need to have
$\frac{1}{1+\alpha}=2Q$ where $Q$ is a fractional number in the
range of $1/4<Q<1/2$ and its denominator should be odd. Here also
the EoS parameter of the modified variable Chaplygin $f(R)$-gravity
model cannot accommodate the transition from $\omega_R>-1$ to
$\omega_R<-1$.
\section{$f(R)$ reconstruction in de Sitter space}

The scale factor in de sitter space is defined as
\begin{equation}
a(t)=a_0e^{Ht},~~~H={\rm constant},\label{infaltion a}
\end{equation}
which can describe the early time inflation of the universe
\cite{Nojiri}. Using Eqs. (\ref{R}) and (\ref{infaltion a}) one can
obtain
\begin{equation}
H=\Big(\frac{R}{12}\Big)^{1/2}.\label{H inf}
\end{equation}
Then Eqs. (\ref{density}) and (\ref{pressure}) take the forms
\begin{eqnarray}
k^2\rho_{R}&=& -\frac{1}{2}f(R)+3 H^2f'(R),\label{density i}\\
k^2p_{R}&=& \frac{1}{2}f(R)-3H^2f'(R).
\end{eqnarray}
Also the EoS parameter yields
\begin{equation}
\omega_R=\frac{p_{R}}{\rho_{R}}=-1,\label{wde}
\end{equation}
which behaves like the cosmological constant.

In what follows, we reconstruct different $f(R)$-gravities according
to the polytropic, SCG, GCG, MCG and MVCG models in de Sitter space.
\subsection{Polytropic gas model}

Using the EoS of the polytropic gas model (\ref{pol1}) and
(\ref{wde}) one can get
\begin{equation}\label{Poly eos2}
K{\rho_{R}}^{\frac{1}{n}}=1.
\end{equation}
Substituting Eqs. (\ref{H inf}) and (\ref{Poly eos2}) into
(\ref{density i}) gives
\begin{equation}\label{Poly Equ i}
R f'(R) -2 f(R)-4 k^2\Big(\frac{1}{K}\Big)^n =0.
\end{equation}
Solving the above differential equation yields
\begin{equation}\label{Poly f i}
f(R)=\lambda R^2-2 k^2\Big(\frac{1}{K}\Big)^n,
\end{equation}
where $\lambda$ is an integration constant. Note that in order to
generate the inflation at the early universe as in Starobinsky's
model \cite{Nojiri,Starobinsky}, one may require
\begin{equation}
\lim_{R\rightarrow\infty} f(R)\propto R^2.\label{inf}
\end{equation}
We see that the polytropic $f(R)$-gravity model (\ref{Poly f i}) can
satisfy the requirement of having to inflation (\ref{inf}).
\subsection{SCG model}

For the SCG model, using Eqs. (\ref{SCGEoS}) and (\ref{wde}) one can
obtain
\begin{equation}\label{SCG eos2}
  {\rho_{R}}^{2}=A.
\end{equation}
Inserting Eqs. (\ref{H inf}) and (\ref{SCG eos2}) into (\ref{density
i}) gives
\begin{equation}\label{SCG Equ i}
R f'(R) -2 f(R)-4 k^2\sqrt{A} =0.
\end{equation}
The resulting $f(R)$ is
\begin{equation}\label{SCG f i}
f(R)=\lambda R^2-2 k^2\sqrt{A},
\end{equation}
where $\lambda$ is an integration constant. Note that the standard
Chaplygin $f(R)$-gravity model (\ref{SCG f i}), like polytropic gas
model (\ref{Poly f i}), satisfies the inflation condition
(\ref{inf}).
\subsection{GCG model}

Regarding the GCG model, with the help of Eqs. (\ref{GCGEoS}) and
(\ref{wde}) one can get
\begin{equation}\label{GCG eos2}
  \rho_{R}^{\alpha+1}=A.
\end{equation}
Replacing Eqs. (\ref{H inf}) and (\ref{GCG eos2}) into (\ref{density
i}) gives
\begin{equation}\label{GCG Equ i}
R f'(R) -2 f(R)-4 k^2{A}^{\frac{1}{\alpha+1}} =0.
\end{equation}
This gives
\begin{equation}\label{GCG f i}
f(R)=\lambda R^2-2 k^2{A}^{\frac{1}{\alpha+1}},
\end{equation}
where $\lambda$ is an integration constant. Here the inflation
condition (\ref{inf}) is also held for the generalized Chaplygin
$f(R)$-gravity model.
\subsection{MCG model}

Substituting the EoS of the MCG model (\ref{MCGEoS}) into
(\ref{wde}) reduces to
\begin{equation}\label{MCG eos2}
  \rho_{R}^{\alpha+1}=\frac{A}{1+B}.
\end{equation}
With the help of Eqs. (\ref{H inf}) and (\ref{MCG eos2}), the
differential Eq. (\ref{density i}) yields
\begin{equation}\label{MCG Equ i}
R f'(R) -2 f(R)-4 k^2\Big(\frac{A}{1+B}\Big)^{\frac{1}{\alpha+1}}
=0.
\end{equation}
The resulting modified Chaplygin $f(R)$-gravity model is obtained as
\begin{equation}\label{MCG f i}
f(R)=\lambda R^2-2 k^2\Big(\frac{A}{1+B}\Big)^{\frac{1}{\alpha+1}},
\end{equation}
where $\lambda$ is an integration constant and the term $\lambda
R^2$ confirms the inflation condition (\ref{inf}).
\subsection{MVCG Model}

For the MVCG model, from Eqs. (\ref{MVCGEoS}) and (\ref{wde}) one
can find
\begin{equation}\label{MVCG eos2}
  a^n\rho_{R}^{\alpha+1}=\frac{A}{1+B}.
\end{equation}
Using Eqs. (\ref{MVCG}) and (\ref{MVCG eos2}) one can obtain
\begin{equation}\label{MVCG eos33}
  \frac{1}{a^n}=\Big(\frac{-nA}{C(1+B)\beta}\Big)
  ^{\frac{n}{\beta}}.
  \end{equation}
Substituting Eq. (\ref{MVCG eos33}) into (\ref{MVCG eos2}) yields
\begin{equation}\label{MVCG eos3}
  {\rho_{R}}=\left[\Big(\frac{A}{1+B}\Big)\Big(\frac{-n A}{C(1+B)\beta}\Big)^{\frac{n}{\beta}}
  \right]^{\frac{1}{\alpha+1}}.
\end{equation}
Putting Eqs. (\ref{H inf}) and (\ref{MVCG eos3}) into (\ref{density
i}) yields
\begin{equation}\label{MVCG Equ i}
R f'(R) -2 f(R)-4 k^2\left[\Big(\frac{A}{1+B}\Big)\Big(\frac{-n A}{C
(1+B)\beta}\Big)^{\frac{n}{\beta}}\right]^{\frac{1}{\alpha+1}} =0.
\end{equation}
This gives
\begin{equation}\label{MVCG f i}
f(R)=\lambda R^2-2 k^2\left[\Big(\frac{A}{1+B}\Big)\Big(\frac{-n
A}{C (1+B)
\beta}\Big)^{\frac{n}{\beta}}\right]^{\frac{1}{\alpha+1}},
\end{equation}
where $\lambda$ is an integration constant. Here the inflation
condition (\ref{inf}) is also satisfied.

In summary one can conclude that in de Sitter space, the obtained
$f(R)$-gravity models corresponding to the polytropic, SCG, GCG, MCG
and MVCG models take the form
\begin{equation}\label{form final}
    f(R)=\lambda_1 R^2+\lambda_2,
\end{equation}
where $\lambda_1$ is the integration constant and $\lambda_2$ is a
constant which depends on the model parameters.
\section{Conclusions}

Here, we investigated the polytropic gas, SCG, GCG, MCG and MVCG
models of DE in the framework of $f(R)$-gravity. Among other
approaches related with a variety of
 DE models, a very promising approach to DE is related with the modified
theories of gravity known as $f(R)$-gravity, in which DE emerges
from the modification of geometry. We reconstructed different
theories of modified gravity based on the $f(R)$ action in the
spatially-flat FRW universe and according to the polytropic gas and
different versions of the Chaplygin gas DE scenarios. We assumed two
classes of scale factors containing i) $a=a_0(t_s-t)^{-h}$ and ii)
$a=a_0e^{Ht}$ which can describe the present accelerating expansion
and the early time inflation of the universe, respectively.
Furthermore, we obtained the EoS parameters of the corresponding
$f(R)$-gravity models. Our calculations show that for the first
class of scale factors, the EoS parameters of the polytropic, SCG,
GCG, MCG and MVCG $f(R)$-gravities can behave like phantom or
quintessence DE models. Whereas the EoS parameters of the
above-mentioned models cannot accommodate the transition from the
quintessence state, $\omega_R>-1$, to the phantom regime,
$\omega_R<-1$, as indicated by recent observations. In the other
words, they cannot behave like the quintom model \cite{YFCai,YFCai1}
in which the EoS parameter is able to evolve across the cosmological
constant boundary. For the second class of scale factors, the EoS
parameter behaves like the cosmological constant. Also the
$f(R)$-gravities in de Sitter space corresponding to the polytropic
gas, SCG, GCG, MCG and MVCG models can predict the early time
inflation of the universe.

Finally, it is worth noting that a viable model of explaining
current acceleration of our universe should take into account the
perturbation analysis which can help to check the stability of the
model. In a full relativistic treatment to discuss the stability of
our models under linear perturbation, we need to define gauge
invariant variables and solve perturbed Einstein's equation and
conservation equations. However, it is beyond the scope of the
present work.
\section*{Acknowledgements}
The authors thank the anonymous referee for a number of valuable
suggestions. The work of K. Karami has been supported financially by
Research Institute for Astronomy and Astrophysics of Maragha (RIAAM)
under research project No. 1/2064.



\begin{thebibliography}{}

\bibitem{riess} A.G. Riess, et al., Astron. J. {\bf 116}, 1009 (1998).

\bibitem{perl} S. Perlmutter, et al., Astrophys. J. {\bf 517}, 565 (1999).

\bibitem{Padmanabhan} T. Padmanabhan, Phys. Rep. {\bf 380}, 235 (2003).

\bibitem{Peebles} P.J.E. Peebles, B. Ratra, Rev. Mod. Phys. {\bf 75}, 559 (2003).

\bibitem{Copeland} E.J. Copeland, M. Sami, S. Tsujikawa, Int. J. Mod. Phys. D {\bf 15},
1753 (2006).

\bibitem{Karami2} K. Karami, S. Ghaffari, J. Fehri, Eur. Phys. J. C {\bf 64}, 85
(2009).

\bibitem{Kamench} A. Kamenshchik, U. Moschella, V. Pasquier, Phys. Lett. B {\bf 487},
7 (2000).

\bibitem{Kamenshchik} A. Kamenshchik, U. Moschella, V. Pasquier, Phys. Lett. B {\bf 511},
265 (2001).

\bibitem{Bilic} N. Bilic, G.B. Tupper, R.D. Viollier, Phys. Lett. B {\bf 535}, 17
(2002).

\bibitem{Wu} P. Wu, H. Yu, Astrophys. J. {\bf 658}, 663 (2007).

\bibitem{Lu} J. Lu, Y. Gui, L.X. Xu, Eur. Phys. J. C {\bf 63}, 349 (2009).

\bibitem{sandvik} H.B. Sandvik, et al., Phys. Rev. D {\bf 69},
123524 (2004).

\bibitem{Bean} R. Bean, O. Dor\'{e}, Phys. Rev. D {\bf 68}, 023515 (2003).

\bibitem{Bento} M.C. Bento, O. Bertolami, A.A. Sen, Phys. Rev. D {\bf
66}, 043507 (2002).

\bibitem{Gorini} V. Gorini, A. Kamenshchik, U. Moschella, Phys. Rev. D {\bf 67},
063509 (2003).

\bibitem{Alam} U. Alam, et al., Mon. Not. R. Astron. Soc. {\bf 344}, 1057 (2003).

\bibitem{Bento1} M.C. Bento, O. Bertolami, A.A. Sen, Phys. Rev. D {\bf 70}, 083519
(2004).

\bibitem{Jamil} M. Jamil, M.A. Rashid, Eur. Phys. J. C {\bf 56}, 429 (2008).

\bibitem{Karami} K. Karami, S. Ghaffari, M.M. Soltanzadeh, Astrophys. Space Sci. {\bf
331}, 309 (2011).

\bibitem{Benaoum} H.B. Benaoum, hep-th/0205140.

\bibitem{Debnath} U. Debnath, A. Banerjee, S. Chakraborty, Class. Quantum Grav. {\bf
21}, 5609 (2004).

\bibitem{Jamil2} M. Jamil, M.A. Rashid, Eur. Phys. J. C {\bf 60}, 141 (2009).

\bibitem{Anup} A.K. Singha, U. Debnath, Int. J. Mod. Phys. D {\bf 16}, 117 (2007).

\bibitem{Jamil3} M. Jamil, Astrophys. Space Sci. {\bf 312}, 295 (2007).

\bibitem{Jamil4} M. Jamil, M.A. Rashid, Eur. Phys. J. C {\bf 58}, 111 (2008).

\bibitem{Malekjani} M. Malekjani, A. Khodam-Mohammadi, Int. J. Mod. Phys. D {\bf 20},
281 (2011).

\bibitem{Tekola} A.G. Tekola, arXiv:0706.0804.

\bibitem{Capozziello1} S. Capozziello, Int. J. Mod. Phys. D {\bf 11}, 483 (2002).

\bibitem{Sobouti} Y. Sobouti, Astron. Astrophys. {\bf 464}, 921
(2007).

\bibitem{NojiriOdin} S. Nojiri, S.D. Odintsov, Phys. Rev. D {\bf 74}, 086005 (2006).

\bibitem{Nojiri1} S. Nojiri, S.D. Odintsov, J. Phys. Conf. Ser. {\bf 66}, 012005
(2007).

\bibitem{Nojiri2} S. Nojiri, S.D. Odintsov, D. Saez-Gomez, arXiv:0908.1269.

\bibitem{Nojiri3} S. Nojiri, S.D. Odintsov, Phys. Rept. {\bf 505}, 59 (2011).

\bibitem{husawiski} A.A. Starobinsky, Phys. Lett. B {\bf 91}, 99 (1980).

\bibitem{Kerner} R. Kerner, Gen. Relativ. Gravit. {\bf 14}, 453 (1982).

\bibitem{Barrow} J. Barrow, A. Ottewill, J. Phys. A {\bf 16}, 2757 (1983).

\bibitem{Faraoni} V. Faraoni, Phys. Rev. D {\bf 74}, 023529 (2006).

\bibitem{Schmidt} H.J. Schmidt, Int. J. Geom. Math. Phys. {\bf 4}, 209 (2007).

\bibitem{Hu} W. Hu, I. Sawicki, Phys. Rev. D {\bf 76}, 064004 (2007).

\bibitem{noj abdalla} S. Nojiri, S.D. Odintsov, Gen. Relativ. Gravit. {\bf 36},
1765 (2004).

\bibitem{Odintsov} S.D. Odintsov, S. Nojiri, Mod. Phys. Lett. A {\bf 19}, 627 (2004).

\bibitem{Abdalla} M.C.B. Abdalla, S. Nojiri, S.D. Odintsov, Class. Quantum Grav. {\bf
22}, L35 (2005).

\bibitem{Noj2} S. Nojiri, S.D. Odintsov, Phys. Lett. B {\bf 576}, 5
(2003).

\bibitem{Carroll} S.M. Carroll, et al., Phys. Rev. D {\bf 70}, 043528 (2004).

\bibitem{Capozziello} S. Capozziello, S. Nojiri, S.D.
Odintsov, Phys. Lett. B {\bf 634}, 93 (2006).

\bibitem{Nojiri5} S. Nojiri, S.D. Odintsov, D. S\'{a}ez-G\'{o}mez, Phys. Lett. B {\bf
681}, 74 (2009).

\bibitem{Bisabr} Y. Bisabr, Phys. Scr. {\bf 80}, 045902 (2009).

\bibitem{Karami4} K. Karami, M.S. Khaledian, JHEP {\bf 03}, 086 (2011).

\bibitem{Khodam-Mohammadi} A. Khodam-Mohammadi, P. Majari, M. Malekjani, Astrophys. Space Sci.
{\bf 331}, 673 (2011).

\bibitem{bengochea} G.R. Bengochea, R. Ferraro, Phys. Rev. D {\bf 79}, 124019 (2009).

\bibitem{YFCai} Y.F. Cai, et al., Phys. Lett. B {\bf 646}, 141 (2007).

\bibitem{YFCai1} Y.F. Cai, et al., Phys. Rep. {\bf 493}, 1 (2010).

\bibitem{JQXia} J.Q. Xia, et al., Int. J. Mod. Phys. D {\bf 17}, 1229
(2008).

\bibitem{NojiriOdin2009} S. Nojiri, S.D. Odintsov, arXiv:0910.1464.

\bibitem{Nozari} K. Nozari, T. Azizi, Phys. Lett. B {\bf 680}, 205 (2009).

\bibitem{Nojiri} S. Nojiri, S.D. Odintsov, Int. J. Geom. Meth. Mod. Phys. {\bf 4},
115 (2007).

\bibitem{Sadjadi} H.M. Sadjadi, Phys. Rev. D {\bf 73}, 063525 (2006).

\bibitem{NojiriOdin1} S. Nojiri, S.D. Odintsov, Phys. Lett. B {\bf 657}, 238
(2007).

\bibitem{Nojiri8} S. Nojiri, S.D. Odintsov, Phys. Rev. D {\bf 77}, 026007 (2008).

\bibitem{Sahni} U. Alam, V. Sahni, A.A. Starobinsky, JCAP {\bf06}, 008 (2004).

\bibitem{Huterer5} D. Huterer, A. Cooray, Phys. Rev. D {\bf 71}, 023506 (2005).

\bibitem{Wang6} Y. Wang, M. Tegmark, Phys. Rev. D {\bf 71}, 103513 (2005).

\bibitem{YFCai2} Y.F. Cai, J. Wang, Class. Quantum Grav. {\bf 25}, 165014
(2008).

\bibitem{Starobinsky} A.A. Starobinsky, JETP Lett. {\bf 86}, 157 (2007).

\end{thebibliography}
\end{document}